\documentclass[fleqn,usenatbib,letters]{mnras}

\usepackage[T1]{fontenc}
\DeclareRobustCommand{\VAN}[3]{#2}
\let\VANthebibliography\thebibliography
\def\thebibliography{\DeclareRobustCommand{\VAN}[3]{##3}\VANthebibliography}

\usepackage{graphicx}	% Including figure files
\usepackage{amsmath}	% Advanced maths commands
\usepackage{amssymb}	% Extra maths symbols

\usepackage{bm}

\usepackage{tikz,xcolor,hyperref}
\usepackage{graphicx}
\graphicspath{{images/}}

\definecolor{lime}{HTML}{A6CE39}
\DeclareRobustCommand{\orcidicon}{%
	\begin{tikzpicture}
	\draw[lime, fill=lime] (0,0) 
	circle [radius=0.16] 
	node[white] {{\fontfamily{qag}\selectfont \tiny ID}};
	\draw[white, fill=white] (-0.0625,0.095) 
	circle [radius=0.007];
	\end{tikzpicture}
	\hspace{-2mm}
}

\newcommand{\orcidVP}{\href{https://orcid.org/0000-0002-3031-062X}{\orcidicon}}
\newcommand{\orcidEV}{\href{https://orcid.org/0000-0003-2742-6872}{\orcidicon}}

\newcommand{\Msun}{M_\odot}

\newcommand{\trh}[1][]{t_\mathrm{rh#1}}
\newcommand{\tcc}[1][]{t_\mathrm{cc#1}}
\newcommand{\ra}[1][]{r_\mathrm{a#1}}
\newcommand{\rh}[1][]{r_\mathrm{h#1}}

\newcommand{\disp}[1][]{\sigma_\mathrm{#1}}
\newcommand{\meq}[1][]{m_\mathrm{eq#1}}
\newcommand{\etam}[1][]{\eta_\mathrm{#1}}

\usepackage{newtxtext,newtxmath}

\title[Star cluster evolution towards energy equipartition]{New insights into star cluster evolution towards energy equipartition}

\author[Pavl\'ik \& Vesperini]{
V\'aclav Pavl\'ik\,$^{1}$\thanks{E-mail: vpavlik@iu.edu}\orcidVP,
Enrico Vesperini\,$^{1}$\orcidEV
\\
$^{1}$Indiana University, Department of Astronomy, Swain Hall West, 727 E 3$^\text{rd}$ Street, Bloomington, IN, 47405, USA
}

\date{Accepted 2021 March 10. Received 2021 March 5; in original form 2021 February 14}

\pubyear{2021}

\begin{document}
\label{firstpage}
\pagerange{\pageref{firstpage}--\pageref{lastpage}}
\maketitle

% Abstract of the paper, not more than 250 words (200 words for Letters)
\begin{abstract}
We present the results of a study aimed at exploring the evolution towards energy equipartition in star cluster models with different initial degrees of anisotropy in the velocity distribution.
Our study reveals a number of novel aspects of the cluster dynamics and shows that the rate of evolution towards energy equipartition (1) depends on the initial degree of radial
velocity anisotropy -- it is more rapid for more radially anisotropic systems; and (2) differs for the radial and the tangential components of the velocity dispersion. (3) The outermost regions of the initially isotropic system evolve towards a state of `inverted' energy equipartition in which high-mass stars have a larger velocity dispersion than low-mass stars -- this inversion originates from the mass-dependence of the tangential velocity dispersion whereas
the radial velocity dispersion shows no anomaly.
Our results add new fundamental elements to the theoretical framework needed to interpret the wealth of recent and upcoming observational studies of stellar kinematics in globular clusters, and shed further light on the link between the clusters' internal kinematics, their formation and evolutionary history.
\end{abstract}

% Select between one and six entries from the list of approved keywords.
% Don't make up new ones.

%%%%%%%%%%%%%%%%%%%%%%%%%%%%%%%%%%%%%%%%%%%%%%%%%%

%%%%%%%%%%%%%%%%% BODY OF PAPER %%%%%%%%%%%%%%%%%%

\section{Introduction}
\label{sec:intro}

The classic picture of the dynamical properties of globular clusters (GCs) has recently been enriched by a number of theoretical and observational studies. They have expanded our understanding of the stellar kinematics in these systems and the link between the GC present-day kinematic properties and evolutionary history. Several observational studies have shown that GCs are characterised by internal kinematics more complex than that assumed in the standard picture according to which these systems are non-rotating and have an isotropic velocity distribution. Large radial velocity surveys and proper motion studies based on {\it HST} and {\it Gaia} observations have shown that GCs instead often display internal rotation \citep[see, e.g.,][]{bellini_hstV,ferraro_etal,MUSE} or
radially anisotropic velocity distribution \citep[see, e.g.,][]{hst_UV_legacy,watkins_hst,jindal_gaia}.
These discoveries spurred new theoretical efforts to understand the role of the internal dynamical processes and the external tidal field of the host galaxy on the observed GC kinematics.

A number of studies \citep[see, e.g.,][]{fokkerplanck_rotI,fokkerplanck_rotII,ernst_etal,hong_etal,Tio_Ves_Var17} have explored the evolution of rotating GCs and have shown that clusters gradually lose memory of their initial rotation as a result of two-body relaxation and the loss of angular momentum carried away by escaping star.
The development and evolution of anisotropy in the velocity distribution has also received significant attention in several theoretical studies which analysed the connection with the cluster's structural properties, the various stages of its dynamical evolution, and the extent of its mass loss \citep[see, e.g.,][and references therein]{giersz_heggie,hurley_shara,Tio_Ves_Var16}

Recently, {\it HST} proper motion data enabled exploring the stellar mass dependence of the velocity dispersion in a few GCs  and to carry out the first measurements of energy equipartition (EEP), one of the fundamental effects of two-body relaxation
\citep{baldwin_hst,hst_legacy_xviii,libralato_hst}. On the theoretical side, a few studies have revisited the evolution towards EEP in GCs \citep{omegaCen_noequip,webb_vesperini_a} and revealed how it links to the GC dynamical history \citep{bianchini_core_collapse,cohen_GCs,libralato_hst}.

In this Letter, we present the results of a study aimed at further exploring a number of fundamental dynamical aspects of the evolution towards EEP in star cluster models.
In particular, we focus on
(1) the dependence of the evolution towards EEP on the extent of the initial radial anisotropy in the velocity dispersion,
(2) the degree of EEP for radial and tangential velocity dispersion, and
(3) the variation of the degree of EEP with the clustercentric distance.
Our investigation shows novel fundamental results concerning all these three aspects and adds new elements to the theoretical framework needed to interpret the wealth of recent and upcoming observational data on the internal kinematics of GCs.

\section{Methods}
\label{sec:methods}
We ran a set of $N$-body models of star clusters with $N_0 = 10^5$ stars, initially distributed according to a \citet{king_model} profile, with the central dimensionless potential $W_0 = 6$. Stellar masses were drawn from the \citet{kroupa} initial mass function in the range $0.1\,\Msun \leq m \leq 1.0\,\Msun$.
We did not include any primordial binaries but the formation and evolution of dynamical binaries was allowed. For the purpose of our subsequent analysis, dynamically formed binary stars were, however, treated as single objects with positions and velocities equal to that of the binary's centre of mass.
The models were integrated with \textsc{nbody6++gpu} \citep{nbody6pp}; since our study was focused on the effects of two-body relaxation, we did not include stellar evolution. 
The modelled clusters were placed on a circular orbit in a point-like Galactic potential with the initial ratio of the tidal radius, $r_t$, to the King model truncation radius equal to ten.

\begin{figure}
	\centering
	\includegraphics[width=\linewidth]{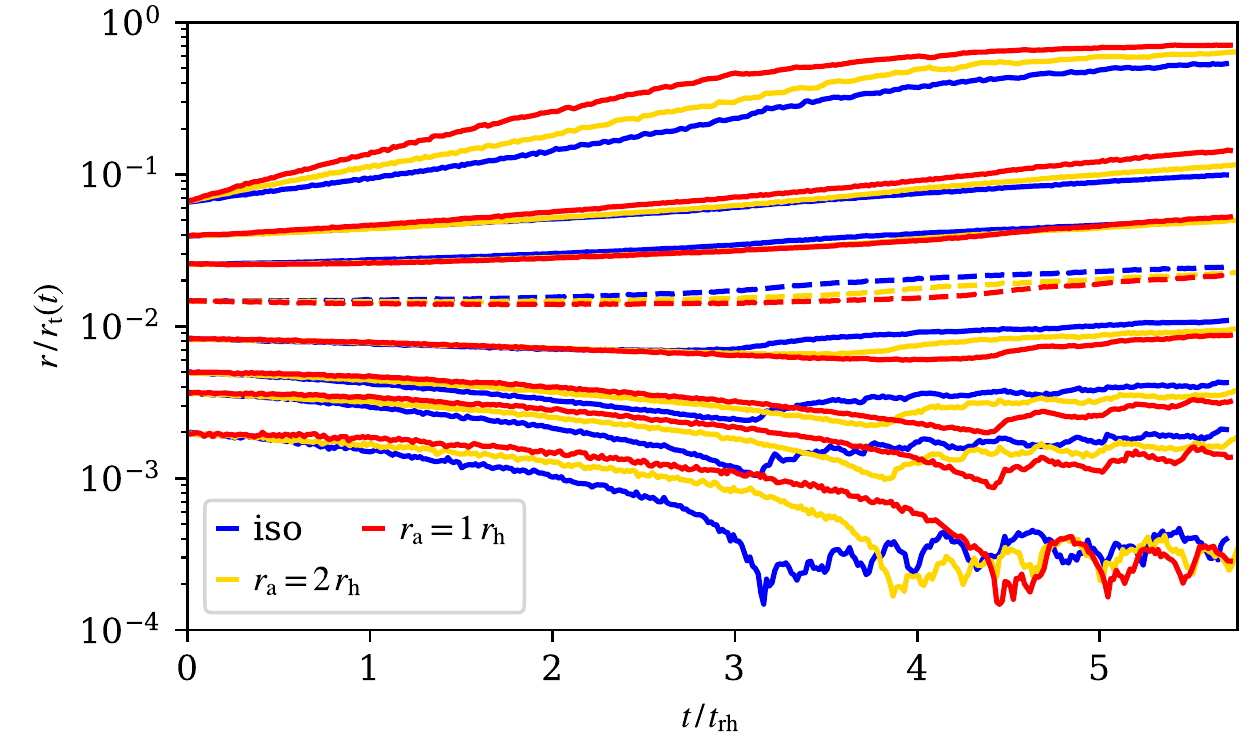}
	\caption{Time evolution of the 1, 5, 10, 25, 50, 75, 90 and 99\,\% Lagrangian radii (normalised to the tidal radius). Time is in the units of the initial half-mass relaxation time.}
	\label{fig:lagr}
\end{figure}

We considered three different initial conditions with the same spatial structure but different degrees of radial velocity anisotropy, following the Osipkov--Merritt profile
\begin{equation}
	\label{eq:aniso}
	1 - \frac{\disp[tan]^2}{2 \disp[rad]^2} = \frac{(r/\ra)^2}{1 + (r/\ra)^2}
\end{equation}
(see, e.g., \citealt{binney_tremaine}; numerically set up with the tool \textsc{agama} from \citealt{agama}),
where $\disp[rad]$ and $\disp[tan]$ are the radial and tangential velocity dispersions, respectively, and the anisotropy radius, $\ra$, is a parameter that marks the boundary between the inner isotropic region and the outer regions where the velocity distribution becomes increasingly radially anisotropic.
We ran a single realisation for each initial configuration -- one model was initially isotropic ($\ra{\rightarrow}\infty$) and the other two had $\ra = \rh[,0]$ and $\ra=2\,\rh[,0]$ (where $\rh[,0]$ is the initial half-mass radius of the cluster).
We note that despite the kinematic differences, all three models had the same initial spatial structure and the half-mass relaxation time defined as $\trh = 0.138 N_0\,\rh[,0]^{3/2} \big/ \ln{(0.02 N_0)}$, in H\'enon units.

\section{Results}
\label{sec:results}

Before proceeding to the analysis of the evolution towards EEP and its dependence on the initial radial velocity anisotropy, it is instructive to first compare the general structural evolution of the three models considered in our study.

\subsection{Structural evolution}

As we show in Fig.~\ref{fig:lagr}, the evolution of the modelled clusters depends on their internal kinematics. The time needed to reach core collapse, $\tcc$ \citep[estimated as in][]{pavl_subr}, increases with the extent of the initial radial velocity anisotropy; specifically, $\tcc[,i] \approx 3.2\,\trh$ in the isotropic model, $\tcc[,a2] \approx 3.8\,\trh$ and $\tcc[,a1] \approx 4.4\,\trh$ in the models with $\ra = 2\,\rh[,0]$ and $1\,\rh[,0]$, respectively.
The main features of the evolution of the inner Lagrangian radii -- such as the depth of core collapse and the post-core-collapse oscillations -- appear to be independent of the initial velocity anisotropy, while in the outer regions more radially anisotropic models tend to fill their tidal radii more rapidly.
The dependence of the core-collapse timescale on the initial radial anisotropy found in our simulations is consistent with that reported by \citet{breen_var_heg} in a study of equal-mass isolated stellar systems with different degrees of tangential and radial anisotropy. We refer to that paper for further discussion of the possible dynamical processes behind the link between anisotropy and the core-collapse time.

\subsection{Energy equipartition}
\label{sec:equip}

One of the manifestations of the effects of two-body relaxation in GCs is their evolution towards EEP. As a GC evolves, energy exchanges during stellar encounters gradually modify the velocity dispersion of stars with different masses. As a result of the evolution towards EEP, at a given distance from the cluster centre, the velocity dispersion tends to decrease for larger stellar masses. Such evolution of the velocity dispersion is more rapid in the cluster inner regions where the local relaxation timescale is shorter.

EEP in GCs is often quantified by a parameter $\eta$, defined by  the relation $\disp \propto m^{\eta}$. Several theoretical studies \citep[see, e.g.,][]{omegaCen_noequip,webb_vesperini_a} have shown that full EEP (that corresponds to a value of $\eta=-0.5$) is never established even in the core region.

Recently, \citet{bianchini_meq} have shown that the variation of the velocity dispersion with the stellar mass is better described by the exponential function
\begin{equation}
	\label{eq:meq}
	\disp(r,m) \propto \exp{[-m / (2\meq)]} \,.
\end{equation}
In this case the equipartition mass, $\meq$, serves as the parameter quantifying the degree of EEP.
The logarithmic slope $-1/(2 \meq)$ is zero and $\meq$ infinite in a system where $\disp$ does not depend on mass, and $\meq$ decreases as a cluster evolves towards EEP.

\begin{figure*}
	\centering
	\includegraphics[width=\linewidth]{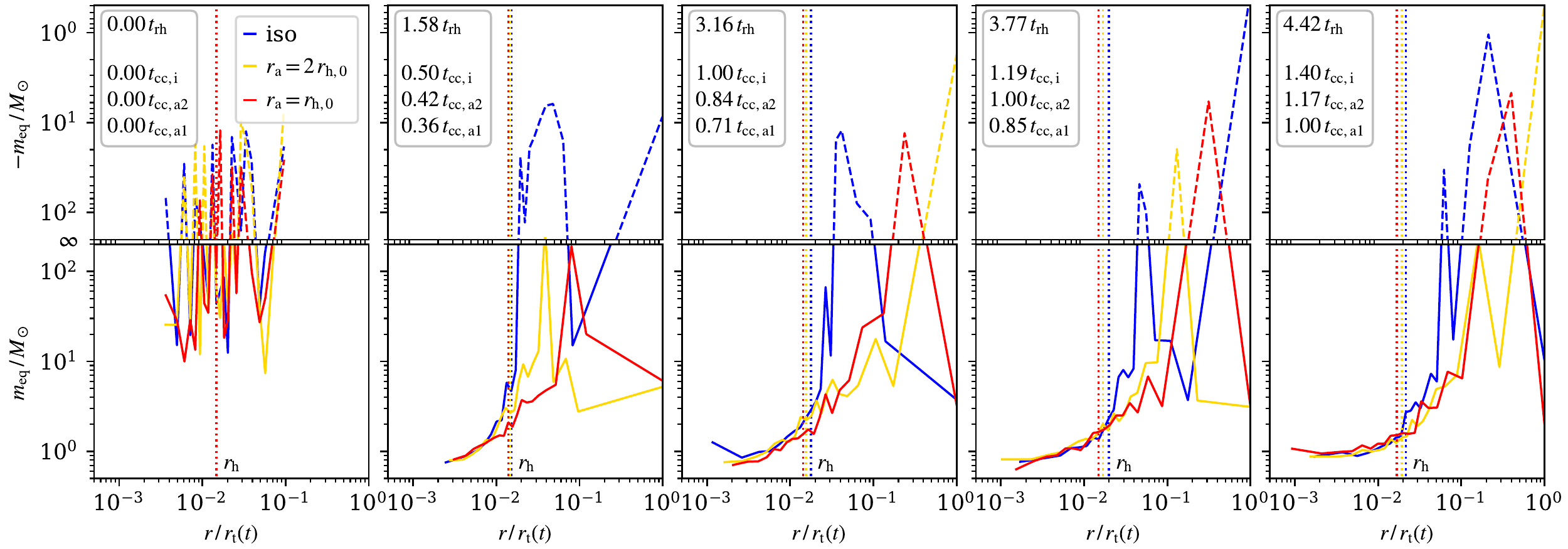}
	\vspace{-10pt}
	\caption{Radial profiles of the equipartition mass. Several snapshots are plotted at times specified in units of the half-mass relaxation time, $\trh$, and the core-collapse time of each model, $\tcc$. The radius is normalised by the tidal radius at a given time. Vertical dotted lines show the half-mass radii, colour-coded for each model as shown in the legend. Each panel is separated vertically between the domains where $\meq > 0$ (bottom, solid lines) and $\meq < 0$ (top, dashed lines). The top vertical axis is inverted to show $-\meq$ correctly, hence, the separation between the top and bottom panels represents $\meq{\rightarrow}\pm\infty$.}
	\label{fig:meq_r_t}
\end{figure*}

In Fig.~\ref{fig:meq_r_t}, we plot the radial profiles of $\meq$ in the modelled clusters for several time snapshots. The left panel shows the initial conditions. Although no initial mass--velocity dependence is prescribed in the models, due to the discrete nature of $N$-body systems, $-1/(2 \meq)$ is not strictly zero but instead oscillates around zero. This yields large values of $\meq$ which are randomly positive or negative.

\begin{figure*}
	\centering
	\includegraphics[width=\linewidth]{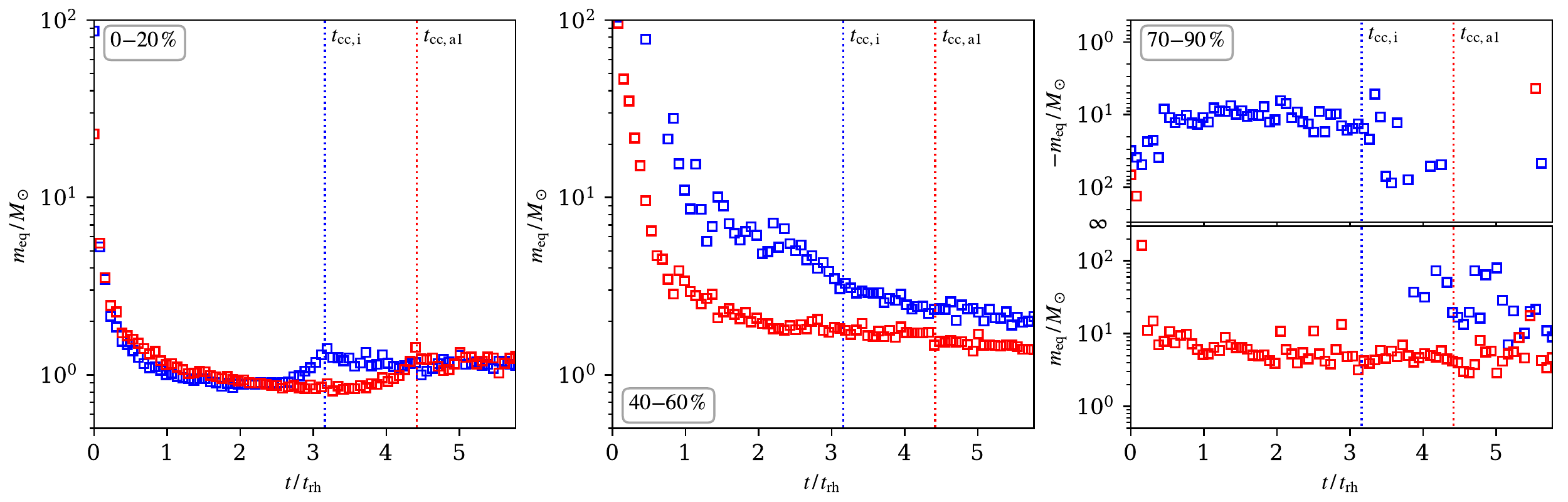}
	\begin{minipage}[b]{.3\linewidth}
		\includegraphics[width=\linewidth,trim=0 180pt 490pt 0,clip]{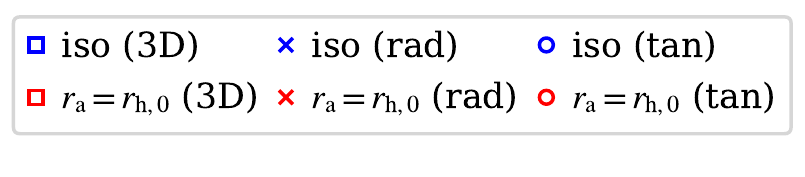}
		\caption{Time evolution of the equipartition mass in our models, calculated from $\disp[rad]$ ($\times$), $\disp[tan]$ ($\bigcirc$) (bottom panels) or the 3D velocity dispersion ($\square$; top panels). Three radial shells are shown, denoted by the Lagrangian radii of a given mass percentage. The outer shell is split into positive (bottom) and negative (top) values of $\meq$. The top vertical axis is inverted to show $-\meq$ correctly -- the separation of both panels represents $\meq{\rightarrow}\pm\infty$ (see Sect.~\ref{sec:equip}).\\}
		\label{fig:meq_t_per}
	\end{minipage}
	\hfill
	\begin{minipage}[b]{.66\linewidth}
		\includegraphics[width=\linewidth,trim=235pt 0 0 0,clip]{{fig_meq_t_RadTan_per_compare_grid_m01-1_w6_100k_0.1}.pdf}
	\end{minipage}
\end{figure*}

The evolution of the central regions of the modelled clusters towards EEP is the most rapid and, consequently, their associated $\meq$ is the smallest. While the system approaches core collapse, $\meq$ gradually decreases in a region extending up to about the half-mass radius.
Our results further show that in the intermediate regions ($\rh \lesssim r \lesssim 3\,\rh$), $\meq$ depends on the initial velocity distribution -- more radially anisotropic models have smaller values of $\meq$ which indicates a faster evolution towards EEP than in the isotropic model. 
At larger distances from the centre the effects of two-body relaxation become less important and as a result the values of $\meq$ increase there.
Some of the outermost shells are characterised by a negative $\meq$ (highlighted in Fig.~\ref{fig:meq_r_t} by dashed lines) corresponding to an `inverted' energy equipartition state in which more massive stars tend to have larger velocity dispersion than low-mass stars.\!\footnote{We emphasise that the values of $-1/(2 \meq)$ where $\meq<0$ are farther from EEP than {\sl any} positive value of $\meq$, and that small negative values are farther from the positive numbers than large negative numbers. Thus, in Figs.~\ref{fig:meq_r_t} \&~\ref{fig:meq_t_per}, we split the panels vertically by a limit $\meq{\rightarrow}\pm\infty$ and invert the top vertical axis with $-\meq$ to show the relationship with EEP correctly.}

\begin{figure*}
	\centering
	\includegraphics[width=\linewidth,trim=0 18pt 0 0,clip]{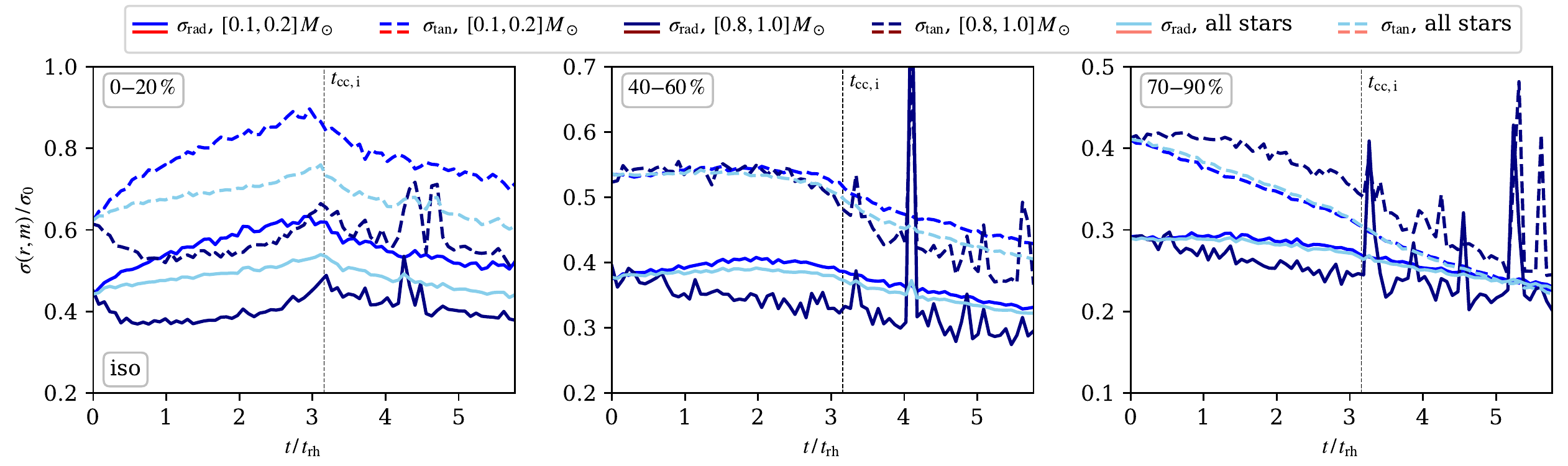}\\
	\includegraphics[width=\linewidth,trim=0 0 0 27pt,clip]{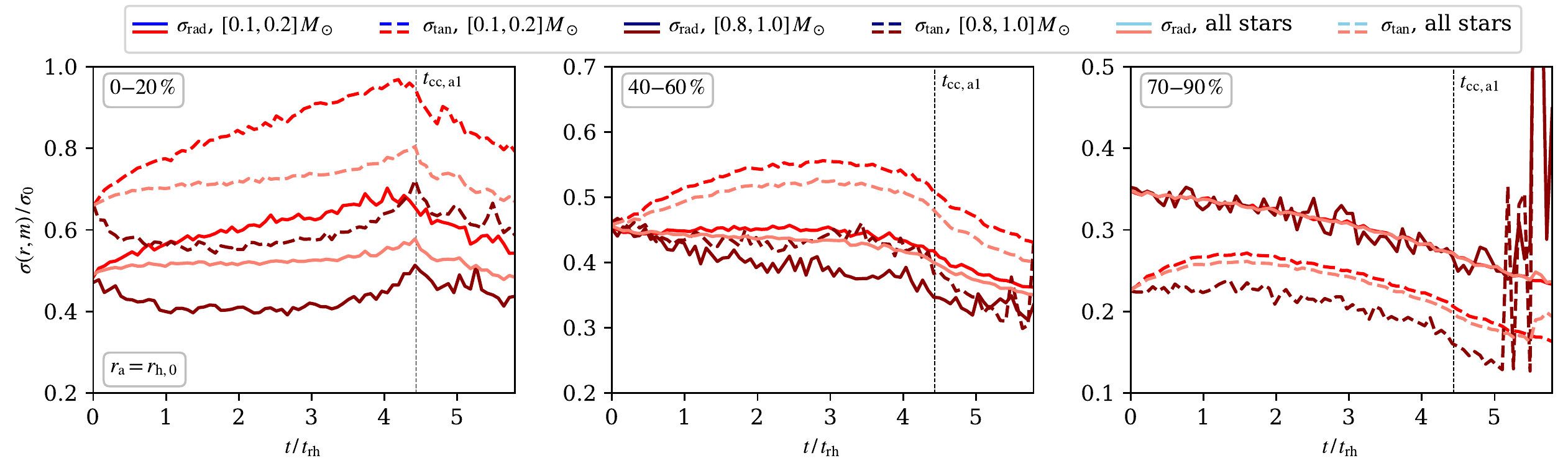}
	\caption{Time evolution of the radial (solid lines) and the tangential components (dashed lines) of the velocity dispersion in three radial shells denoted by the Lagrangian radii of a given percentage, and three mass bins (colour coded) for the isotropic (top rows) and the $\ra=\rh[,0]$ model (bottom rows). The vertical axis is normalised by $\disp[0] = \rh[,0]^{-1/2}$, in H\'enon units.}
	\label{fig:disp_t_per}
\end{figure*}

These results are further supported by the time evolution of $\meq$ shown in Fig.~\ref{fig:meq_t_per}, and $\eta$ in Fig.~\ref{fig:eta_t_per}. Only the isotropic model and the most radially anisotropic model (with $\ra=\rh[,0]$) are compared. In the top panels of both figures, we plot the corresponding parameter fitted to the three-dimensional velocity dispersion.
In both models, the change of $\meq$ and $\eta$ in time is similar in the innermost regions (top left panels), i.e., they approach EEP as expected. We can even see a kinematic trace of core collapse as a local maximum in $\meq$ and $\eta$ at $\tcc$ \citep[cf.][]{bianchini_core_collapse}, which also demonstrates that their method and that of \citet{pavl_subr}, used here to determine $\tcc$, are compatible even in these kinematically distinct clusters.
The top middle and top right panels show the evolution of $\meq$ and $\eta$ in the intermediate and outer shells, and demonstrate the differences between the rate of evolution towards EEP in both models that we discussed above. In addition, the outer shell of the isotropic model clearly shows the presence of an `inverted' EEP: the corresponding values of $\meq[,i]$ are negative during most of the cluster evolution, and the values of $\etam[i]$ remain positive with a slight growth.

In the two bottom panels of Figs.~\ref{fig:meq_t_per} \&~\ref{fig:eta_t_per}, we further explore the evolution towards EEP in the intermediate and the outer shells by plotting the values of $\meq$ and $\eta$, calculated separately from the radial and tangential components of the velocity dispersion. Hereafter, we limit the discussion on $\meq$ but all features and dynamical differences identified from it also apply to $\eta$.
We see that $\meq[,rad]$ differs from $\meq[,tan]$ which indicates that the evolution towards EEP is not isotropic and proceeds at different paces for the radial and the tangential velocity dispersion.
In addition, we also observe a contrasting behaviour of $\meq[,tan]$. In the intermediate shell, the fact that $\meq[,a1,tan] < \meq[,a1,rad]$ keeps the overall equipartition mass of the anisotropic model low (as in Fig.~\ref{fig:meq_r_t}). Instead, in the isotropic model, $\meq[,i,tan]$ maintains large values and is responsible for the slower decrease of the overall $\meq[,i]$\,; only after core collapse do we find $\meq[,i,tan] \approx \meq[,i,rad]$.
The difference is even greater in the outer shell where $\meq[,i,tan]$ of the isotropic model is systematically negative during the entire evolution. As for the anisotropic model, $\meq[,a1,tan]$ remains at values slightly higher than those found in the intermediate shell, while $\meq[,a1,rad]$ is always greater than $\meq[,a1,tan]$ and randomly oscillates between positive and negative values.

\subsection{Velocity dispersion}

In this section we focus on the evolution of the velocity dispersion and discuss its link with the evolution towards EEP and the various features revealed by our previous analysis.
The time evolution of the velocity dispersion is plotted in Fig.~\ref{fig:disp_t_per} for three stellar mass groups and for the same radial shells used in Figs.~\ref{fig:meq_t_per} \&~\ref{fig:eta_t_per}. This selection can best illustrate the key aspects of the evolution towards EEP in the modelled clusters.
By definition, all models started with an isotropic core, which is visible in the left panels of Fig.~\ref{fig:disp_t_per} as the initial split of the radial and tangential components, corresponding to $\disp[tan] = \sqrt{2}\,\disp[rad]$.
The evolution of the velocity dispersion in the inner regions of both models is similar and both remain approximately isotropic. The outer regions, however, differ. The isotropic cluster (top row) gradually develops a radially anisotropic velocity distribution which is visible especially in the outermost shell (right panel).
In this region, we also notice that the decrease of $\disp[tan]$ of the low-mass stars is steeper than that of the high-mass population, which leads to the `inverted' EEP trend.

In the initially anisotropic model with $\ra=\rh[,0]$ (bottom row of Fig.~\ref{fig:disp_t_per}), the velocity anisotropy nearly vanishes in the region around the half-mass radius before the cluster reaches core collapse, and decreases in the outermost region, in particular, for the low-mass stars.
We note that in this model the overall evolution of the velocity dispersion does not show evidence of the anomalous dependence on the stellar mass that leads to the `inverted' EEP found in the isotropic model. It is also interesting to point out that $\disp[rad]$ of the anisotropic model remains approximately independent of mass in the outermost shell and that only $\disp[tan]$ hints at evidence of evolution towards EEP.

The evolution we see in both models is due to the combined effect of (i) preferential escape of low-mass stars on radial orbit, (ii) the segregation of massive stars towards the cluster central regions and the outward migration of low-mass stars, (iii) the evolution of the cluster internal structure, and (iv) the overall expansion of the cluster towards the tidal radius. The relative importance of these effects will be further explored in a future study.

\section{Conclusions}
\label{sec:concl}
In this work we have studied the evolution towards EEP in models of star clusters with different initial radial velocity anisotropy.
Our analysis has revealed a number of novel fundamental aspects in their dynamical evolution and kinematic properties:
\vspace{-6pt}
\begin{itemize}
	\item  Initially anisotropic systems evolve more rapidly towards EEP than the isotropic one.
	\item The rate of evolution towards EEP in the cluster's intermediate and outer regions is different for the radial and the tangential components of the velocity dispersion. Specifically, when we follow the radial component, this evolution is more rapid in the isotropic model; in contrast, for the tangential component, the evolution proceeds more rapidly in the radially anisotropic models.
	\item The outermost regions of the initially isotropic system evolve towards a state of `inverted' EEP in which high-mass stars acquire larger velocity dispersion than low-mass stars. This inversion is driven by an anomalous behaviour of the tangential component of the velocity dispersion while its radial component behaves as expected. This feature is not visible in the initially anisotropic models.
\end{itemize}
\vspace{-6pt}

A number of observational studies have explored the kinematics of the GC cores and inner regions. Our results further highlight the importance of a complete empirical characterisation of GC kinematics that extends to their outermost regions.
In future work, we will explore a broader range of initial conditions using different structural and kinematic properties and the effect of various external tidal fields.
Our preliminary results indicate that the features presented in this work for tidally underfilling clusters hold also for tidally filling systems. For a more direct comparison with observations future models must also include additional ingredients such as a population of primordial binaries and mass loss due stellar evolution.

\section*{Acknowledgements}

VP thanks Steven Shore for valuable discussion.
This research was supported in part by Lilly Endowment, Inc., through its support for the Indiana University Pervasive Technology Institute. % Carbonate acknowledgement
We thank Mirek Giersz for insightful comments that helped us clarify the presentation of the results in this paper.

\section*{Data availability statement}

The data presented in this article may be shared on reasonable request to the corresponding author.

%%%%%%%%%%%%%%%%%%%% REFERENCES %%%%%%%%%%%%%%%%%%

% The best way to enter references is to use BibTeX:
\bibliographystyle{mnras}
\bibliography{MN-21-0536-L} % if your bibtex file is called example.bib

\begin{thebibliography}{}
\makeatletter
\relax
\def\mn@urlcharsother{\let\do\@makeother \do\$\do\&\do\#\do\^\do\_\do\%\do\~}
\def\mn@doi{\begingroup\mn@urlcharsother \@ifnextchar [ {\mn@doi@}
  {\mn@doi@[]}}
\def\mn@doi@[#1]#2{\def\@tempa{#1}\ifx\@tempa\@empty \href
  {http://dx.doi.org/#2} {doi:#2}\else \href {http://dx.doi.org/#2} {#1}\fi
  \endgroup}
\def\mn@eprint#1#2{\mn@eprint@#1:#2::\@nil}
\def\mn@eprint@arXiv#1{\href {http://arxiv.org/abs/#1} {{\tt arXiv:#1}}}
\def\mn@eprint@dblp#1{\href {http://dblp.uni-trier.de/rec/bibtex/#1.xml}
  {dblp:#1}}
\def\mn@eprint@#1:#2:#3:#4\@nil{\def\@tempa {#1}\def\@tempb {#2}\def\@tempc
  {#3}\ifx \@tempc \@empty \let \@tempc \@tempb \let \@tempb \@tempa \fi \ifx
  \@tempb \@empty \def\@tempb {arXiv}\fi \@ifundefined
  {mn@eprint@\@tempb}{\@tempb:\@tempc}{\expandafter \expandafter \csname
  mn@eprint@\@tempb\endcsname \expandafter{\@tempc}}}

\bibitem[\protect\citeauthoryear{{Baldwin}, {Watkins}, {van der Marel},
  {Bianchini}, {Bellini}  \& {Anderson}}{{Baldwin} et~al.}{2016}]{baldwin_hst}
{Baldwin} A.~T.,  {Watkins} L.~L.,  {van der Marel} R.~P.,  {Bianchini} P.,
  {Bellini} A.,   {Anderson} J.,  2016, \mn@doi [\apj]
  {10.3847/0004-637X/827/1/12}, \href
  {https://ui.adsabs.harvard.edu/abs/2016ApJ...827...12B} {827, 12}

\bibitem[\protect\citeauthoryear{{Bellini} et~al.,}{{Bellini}
  et~al.}{2015}]{hst_UV_legacy}
{Bellini} A.,  et~al., 2015, \mn@doi [\apjl] {10.1088/2041-8205/810/1/L13},
  \href {https://ui.adsabs.harvard.edu/abs/2015ApJ...810L..13B} {810, L13}

\bibitem[\protect\citeauthoryear{{Bellini}, {Bianchini}, {Varri}, {Anderson},
  {Piotto}, {van der Marel}, {Vesperini}  \& {Watkins}}{{Bellini}
  et~al.}{2017}]{bellini_hstV}
{Bellini} A.,  {Bianchini} P.,  {Varri} A.~L.,  {Anderson} J.,  {Piotto} G.,
  {van der Marel} R.~P.,  {Vesperini} E.,   {Watkins} L.~L.,  2017, \mn@doi
  [\apj] {10.3847/1538-4357/aa7c5f}, \href
  {https://ui.adsabs.harvard.edu/abs/2017ApJ...844..167B} {844, 167}

\bibitem[\protect\citeauthoryear{{Bianchini}, {van de Ven}, {Norris},
  {Schinnerer}  \& {Varri}}{{Bianchini} et~al.}{2016}]{bianchini_meq}
{Bianchini} P.,  {van de Ven} G.,  {Norris} M.~A.,  {Schinnerer} E.,   {Varri}
  A.~L.,  2016, \mn@doi [\mnras] {10.1093/mnras/stw552}, \href
  {https://ui.adsabs.harvard.edu/abs/2016MNRAS.458.3644B} {458, 3644}

\bibitem[\protect\citeauthoryear{{Bianchini}, {Webb}, {Sills}  \&
  {Vesperini}}{{Bianchini} et~al.}{2018}]{bianchini_core_collapse}
{Bianchini} P.,  {Webb} J.~J.,  {Sills} A.,   {Vesperini} E.,  2018, \mn@doi
  [\mnras] {10.1093/mnrasl/sly013}, \href
  {https://ui.adsabs.harvard.edu/abs/2018MNRAS.475L..96B} {475, L96}

\bibitem[\protect\citeauthoryear{Binney \& Tremaine}{Binney \&
  Tremaine}{2008}]{binney_tremaine}
Binney J.,  Tremaine S.,  2008, Galactic Dynamics: Second Edition.
Princeton Series in Astrophysics, Princeton University Press

\bibitem[\protect\citeauthoryear{{Breen}, {Varri}  \& {Heggie}}{{Breen}
  et~al.}{2017}]{breen_var_heg}
{Breen} P.~G.,  {Varri} A.~L.,   {Heggie} D.~C.,  2017, \mn@doi [\mnras]
  {10.1093/mnras/stx1750}, \href
  {https://ui.adsabs.harvard.edu/abs/2017MNRAS.471.2778B} {471, 2778}

\bibitem[\protect\citeauthoryear{{Cohen}, {Bellini}, {Libralato}, {Correnti},
  {Brown}  \& {Kalirai}}{{Cohen} et~al.}{2021}]{cohen_GCs}
{Cohen} R.~E.,  {Bellini} A.,  {Libralato} M.,  {Correnti} M.,  {Brown} T.~M.,
   {Kalirai} J.~S.,  2021, \mn@doi [\aj] {10.3847/1538-3881/abd036}, \href
  {https://ui.adsabs.harvard.edu/abs/2021AJ....161...41C} {161, 41}

\bibitem[\protect\citeauthoryear{{Einsel} \& {Spurzem}}{{Einsel} \&
  {Spurzem}}{1999}]{fokkerplanck_rotI}
{Einsel} C.,  {Spurzem} R.,  1999, \mn@doi [\mnras]
  {10.1046/j.1365-8711.1999.02083.x}, \href
  {https://ui.adsabs.harvard.edu/abs/1999MNRAS.302...81E} {302, 81}

\bibitem[\protect\citeauthoryear{{Ernst}, {Glaschke}, {Fiestas}, {Just}  \&
  {Spurzem}}{{Ernst} et~al.}{2007}]{ernst_etal}
{Ernst} A.,  {Glaschke} P.,  {Fiestas} J.,  {Just} A.,   {Spurzem} R.,  2007,
  \mn@doi [\mnras] {10.1111/j.1365-2966.2007.11602.x}, \href
  {https://ui.adsabs.harvard.edu/abs/2007MNRAS.377..465E} {377, 465}

\bibitem[\protect\citeauthoryear{{Ferraro} et~al.,}{{Ferraro}
  et~al.}{2018}]{ferraro_etal}
{Ferraro} F.~R.,  et~al., 2018, \mn@doi [\apj] {10.3847/1538-4357/aabe2f},
  \href {https://ui.adsabs.harvard.edu/abs/2018ApJ...860...50F} {860, 50}

\bibitem[\protect\citeauthoryear{{Giersz} \& {Heggie}}{{Giersz} \&
  {Heggie}}{1997}]{giersz_heggie}
{Giersz} M.,  {Heggie} D.~C.,  1997, \mn@doi [\mnras]
  {10.1093/mnras/286.3.709}, \href
  {https://ui.adsabs.harvard.edu/abs/1997MNRAS.286..709G} {286, 709}

\bibitem[\protect\citeauthoryear{{Hong}, {Kim}, {Lee}  \& {Spurzem}}{{Hong}
  et~al.}{2013}]{hong_etal}
{Hong} J.,  {Kim} E.,  {Lee} H.~M.,   {Spurzem} R.,  2013, \mn@doi [\mnras]
  {10.1093/mnras/stt099}, \href
  {https://ui.adsabs.harvard.edu/abs/2013MNRAS.430.2960H} {430, 2960}

\bibitem[\protect\citeauthoryear{{Hurley} \& {Shara}}{{Hurley} \&
  {Shara}}{2012}]{hurley_shara}
{Hurley} J.~R.,  {Shara} M.~M.,  2012, \mn@doi [\mnras]
  {10.1111/j.1365-2966.2012.21668.x}, \href
  {https://ui.adsabs.harvard.edu/abs/2012MNRAS.425.2872H} {425, 2872}

\bibitem[\protect\citeauthoryear{{Jindal}, {Webb}  \& {Bovy}}{{Jindal}
  et~al.}{2019}]{jindal_gaia}
{Jindal} A.,  {Webb} J.~J.,   {Bovy} J.,  2019, \mn@doi [\mnras]
  {10.1093/mnras/stz1586}, \href
  {https://ui.adsabs.harvard.edu/abs/2019MNRAS.487.3693J} {487, 3693}

\bibitem[\protect\citeauthoryear{{Kamann} et~al.,}{{Kamann}
  et~al.}{2018}]{MUSE}
{Kamann} S.,  et~al., 2018, \mn@doi [\mnras] {10.1093/mnras/stx2719}, \href
  {https://ui.adsabs.harvard.edu/abs/2018MNRAS.473.5591K} {473, 5591}

\bibitem[\protect\citeauthoryear{{Kim}, {Einsel}, {Lee}, {Spurzem}  \&
  {Lee}}{{Kim} et~al.}{2002}]{fokkerplanck_rotII}
{Kim} E.,  {Einsel} C.,  {Lee} H.~M.,  {Spurzem} R.,   {Lee} M.~G.,  2002,
  \mn@doi [\mnras] {10.1046/j.1365-8711.2002.05420.x}, \href
  {https://ui.adsabs.harvard.edu/abs/2002MNRAS.334..310K} {334, 310}

\bibitem[\protect\citeauthoryear{{King}}{{King}}{1966}]{king_model}
{King} I.~R.,  1966, \mn@doi [\aj] {10.1086/109857}, \href
  {https://ui.adsabs.harvard.edu/abs/1966AJ.....71...64K} {71, 64}

\bibitem[\protect\citeauthoryear{{Kroupa}}{{Kroupa}}{2001}]{kroupa}
{Kroupa} P.,  2001, \mn@doi [\mnras] {10.1046/j.1365-8711.2001.04022.x}, \href
  {http://adsabs.harvard.edu/abs/2001MNRAS.322..231K} {322, 231}

\bibitem[\protect\citeauthoryear{{Libralato} et~al.,}{{Libralato}
  et~al.}{2018}]{libralato_hst}
{Libralato} M.,  et~al., 2018, \mn@doi [\apj] {10.3847/1538-4357/aac6c0}, \href
  {https://ui.adsabs.harvard.edu/abs/2018ApJ...861...99L} {861, 99}

\bibitem[\protect\citeauthoryear{{Libralato}, {Bellini}, {Piotto}, {Nardiello},
  {van der Marel}, {Anderson}, {Bedin}  \& {Vesperini}}{{Libralato}
  et~al.}{2019}]{hst_legacy_xviii}
{Libralato} M.,  {Bellini} A.,  {Piotto} G.,  {Nardiello} D.,  {van der Marel}
  R.~P.,  {Anderson} J.,  {Bedin} L.~R.,   {Vesperini} E.,  2019, \mn@doi
  [\apj] {10.3847/1538-4357/ab0551}, \href
  {https://ui.adsabs.harvard.edu/abs/2019ApJ...873..109L} {873, 109}

\bibitem[\protect\citeauthoryear{{Pavl{\'\i}k} \& {{\v{S}}ubr}}{{Pavl{\'\i}k}
  \& {{\v{S}}ubr}}{2018}]{pavl_subr}
{Pavl{\'\i}k} V.,  {{\v{S}}ubr} L.,  2018, \mn@doi [\aap]
  {10.1051/0004-6361/201833854}, \href
  {https://ui.adsabs.harvard.edu/\#abs/2018A&A...620A..70P} {620, A70}

\bibitem[\protect\citeauthoryear{{Tiongco}, {Vesperini}  \& {Varri}}{{Tiongco}
  et~al.}{2016}]{Tio_Ves_Var16}
{Tiongco} M.~A.,  {Vesperini} E.,   {Varri} A.~L.,  2016, \mn@doi [\mnras]
  {10.1093/mnras/stv2574}, \href
  {https://ui.adsabs.harvard.edu/abs/2016MNRAS.455.3693T} {455, 3693}

\bibitem[\protect\citeauthoryear{{Tiongco}, {Vesperini}  \& {Varri}}{{Tiongco}
  et~al.}{2017}]{Tio_Ves_Var17}
{Tiongco} M.~A.,  {Vesperini} E.,   {Varri} A.~L.,  2017, \mn@doi [\mnras]
  {10.1093/mnras/stx853}, \href
  {https://ui.adsabs.harvard.edu/abs/2017MNRAS.469..683T} {469, 683}

\bibitem[\protect\citeauthoryear{{Trenti} \& {van der Marel}}{{Trenti} \& {van
  der Marel}}{2013}]{omegaCen_noequip}
{Trenti} M.,  {van der Marel} R.,  2013, \mn@doi [\mnras]
  {10.1093/mnras/stt1521}, \href
  {https://ui.adsabs.harvard.edu/abs/2013MNRAS.435.3272T} {435, 3272}

\bibitem[\protect\citeauthoryear{{Vasiliev}}{{Vasiliev}}{2019}]{agama}
{Vasiliev} E.,  2019, \mn@doi [\mnras] {10.1093/mnras/sty2672}, \href
  {https://ui.adsabs.harvard.edu/abs/2019MNRAS.482.1525V} {482, 1525}

\bibitem[\protect\citeauthoryear{{Wang}, {Spurzem}, {Aarseth}, {Nitadori},
  {Berczik}, {Kouwenhoven}  \& {Naab}}{{Wang} et~al.}{2015}]{nbody6pp}
{Wang} L.,  {Spurzem} R.,  {Aarseth} S.,  {Nitadori} K.,  {Berczik} P.,
  {Kouwenhoven} M.~B.~N.,   {Naab} T.,  2015, \mn@doi [\mnras]
  {10.1093/mnras/stv817}, \href
  {https://ui.adsabs.harvard.edu/abs/2015MNRAS.450.4070W} {450, 4070}

\bibitem[\protect\citeauthoryear{{Watkins}, {van\,der\,Marel}, {Bellini}  \&
  {Anderson}}{{Watkins} et~al.}{2015}]{watkins_hst}
{Watkins} L.\ L.,  {van\,der\,Marel} R.\ P.,  {Bellini} A.,   {Anderson} J.,
  2015, \mn@doi [\apj] {10.1088/0004-637X/803/1/29}, \href
  {https://ui.adsabs.harvard.edu/abs/2015ApJ...803...29W} {803, 29}

\bibitem[\protect\citeauthoryear{{Webb} \& {Vesperini}}{{Webb} \&
  {Vesperini}}{2016}]{webb_vesperini_a}
{Webb} J.~J.,  {Vesperini} E.,  2016, \mn@doi [\mnras] {10.1093/mnras/stw2186},
  \href {https://ui.adsabs.harvard.edu/abs/2016MNRAS.463.2383W} {463, 2383}

\makeatother
\end{thebibliography}

% Don't change these lines
\bsp	% typesetting comment

% %%%%%%%%%%%%%%%%%%%%%%%%%%%%%%%%%%%%%%%%%%%%%%%%%%
% 
% %%%%%%%%%%%%%%%%% APPENDICES %%%%%%%%%%%%%%%%%%%%%

\clearpage
\appendix
\onecolumn

\section{Additional figures}

% Force figure after the title (overrides the mnras.cls definition)
\makeatletter
\def\fps@figure{htb}
\makeatother
\begin{figure*}
	\centering
	\includegraphics[width=\linewidth]{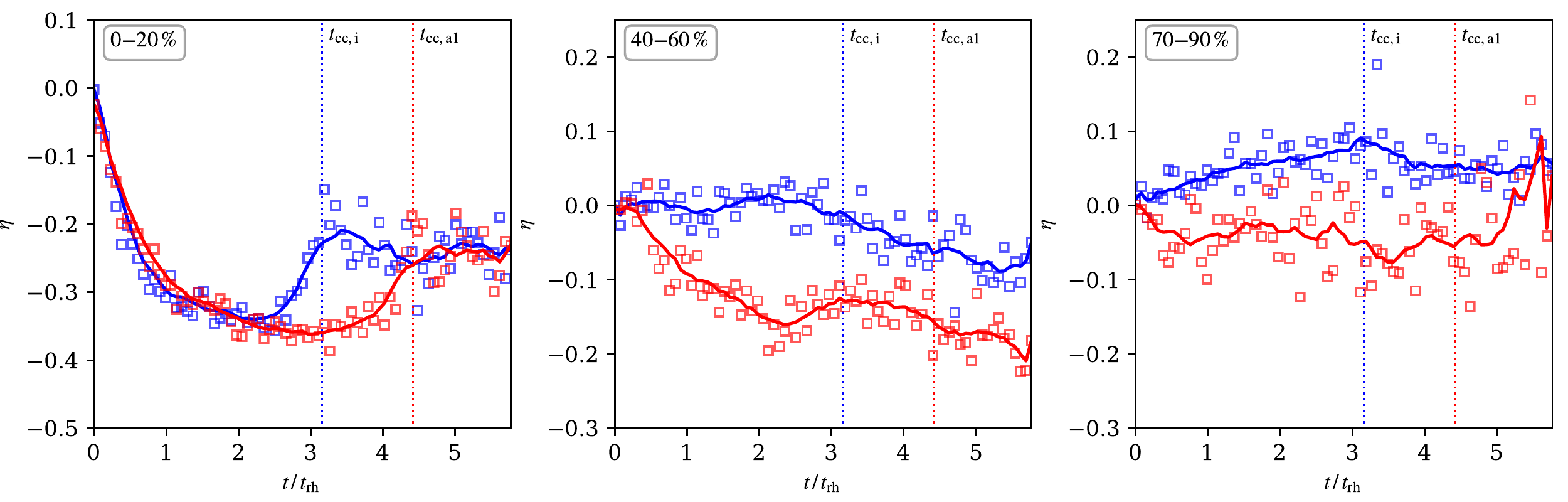}\\
	\begin{minipage}[b]{.3\linewidth}
		\includegraphics[width=\linewidth,trim=0 180pt 490pt 0,clip]{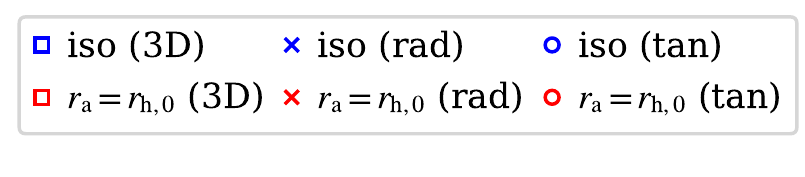}
		\caption{Time evolution of the parameter $\eta$ in our models, calculated from $\disp[rad]$ ($\times$), $\disp[tan]$~($\bigcirc$) (bottom panels)  or the 3D velocity dispersion ($\square$; top panels). Three radial shells are shown, denoted by the Lagrangian radii of a~given mass percentage. The datapoints are calculated at each time snapshot using stars of masses $0.4 \leq m/\Msun \leq 1.0$. Lines representing the simple moving average calculated from the corresponding datapoints are added to guide the eye.}
		\label{fig:eta_t_per}
	\end{minipage}
	\hfill
	\begin{minipage}[b]{.66\linewidth}
		\includegraphics[width=\linewidth,trim=240pt 0 0 0,clip]{{fig_eta_t_RadTan_per_compare_grid_m01-1_w6_100k_0.1}.pdf}
	\end{minipage}
\end{figure*}
% revert the mnras.cls definition
\makeatletter
\def\fps@figure{tbp}
\makeatother

%%%%%%%%%%%%%%%%%%%%%%%%%%%%%%%%%%%%%%%%%%%%%%%%%%

\label{lastpage}
\end{document}